\documentclass[12pt]{article}

\usepackage[document]{ragged2e}
\usepackage{amssymb}
\usepackage{amsmath} 
\usepackage{amsthm}
\usepackage{bm}
\usepackage{graphicx}
\usepackage{float}
\usepackage{hyperref}

\begin{document}

\begin{titlepage}
\begin{center}
\textbf{Modal Smoothing for Analysis of Room Reflections Measured with Spherical Microphone and Loudspeaker Arrays} \\ \vspace{10ex}
Hai Morgenstern$^{1,}$\footnote{e-mail: haimorg@post.bgu.ac.il} and Boaz Rafaely$^1$\\
$^{1}$Department of Electrical and Computer Engineering, \\
Ben Gurion University of the Negev, Beer- Sheva, 84105, Israel\\
%\vspace{10ex}
\end{center}
\end{titlepage}
		
\begin{abstract}
%\thispagestyle{plain}
%\setcounter{page}{2}
%\begin{linenumbers}
\justify
Spatial analysis of room acoustics is an ongoing research topic. 
Microphone arrays have been employed for spatial analyses, with an important objective being the estimation of the direction-of-arrival (DOA) of direct sound and early room reflections using room impulse responses (RIRs).  
An optimal method for DOA estimation is the multiple signal classification algorithm. 
When RIRs are considered, this method typically fails due to the correlation of room reflections, which leads to rank deficiency of the cross-spectrum matrix. 
Preprocessing methods for rank restoration, which may involve averaging over frequency, for example, have been proposed exclusively for spherical arrays. 
However, these methods fail in the case of reflections with equal time delays, which may arise in practice and could be of interest. 
In this paper, a method is proposed for systems that combine a spherical microphone array and a spherical loudspeaker array, referred to as multiple-input multiple-output systems.  
This method, referred to as modal smoothing, exploits the additional spatial diversity for rank restoration and succeeds where previous methods fail, as demonstrated in a simulation study. 
Finally, combining modal smoothing with a preprocessing method is proposed in order to increase the number of DOAs that can be estimated using low-order spherical loudspeaker arrays.
%\end{linenumbers}
%\vspace{50ex}
% word count: 8269
% NEW RUNNING TITLE: Multichannel spatial acoustic analysis
%\vspace{5ex}
%Possible keywords (if required): Microphone arrays, Loudspeaker arrays, spherical arrays, MIMO, spherical harmonics, room acoustics. \\ 
\vspace{5ex}
%Possible PACS (if required): (1) 43.60.Uv - Model-based signal processing; (2) 43.55.Br - Room acoustics: theory and experiment; reverberation, normal modes, diffusion, transient and steady-state response; and, (3) 43.60.Fg - Acoustic array systems and processing, beam-forming. 
				
\end{abstract}
%\addtocounter{page}{3}

%\begin{linenumbers}
\section{INTRODUCTION}
\justify % after each section and subsection
\setlength{\parindent}{5ex} % only for introduction

Spatial analysis of room acoustics has recently gained increased interest in numerous application areas \cite{ribeiro2010using,antonacci2010geometric}.  
In many cases, spatial analysis is facilitated by array processing of measured room impulse responses (RIRs) from a loudspeaker to each microphone in a microphone array \cite{gover2004measurements,khaykin2012acoustic,kuster2004acoustic,gover2002microphone}. 
RIRs may be useful because they facilitate the identification of individual reflections. 

% Problem declaration
In this paper we consider the problem of finding the direction of arrival (DOA) of the direct sound and the room reflections using RIR data measured by multiple microphones. 
% previous methods - beamforming and optimal
Processing methods based on spatial filtering, or beamforming, have been proposed for this task due to their simplicity and robust performance \cite{van2004detection}. 
%Typically, such methods facilitate the control over various measures of array performance, such as white-noise gain and directivity-index \cite{rafaely2005phase}. 
In addition to these methods, optimal methods that are tailored to the specific wave field, such as the Mulitple SIgnal Classification (MUSIC), were reported in the array processing literature \cite{van2004detection,yan2011optimal,sun2011joint,sun2012localization}.
% Shorten from here (MUSIC based methods:
These optimal methods can provide improved spatial analysis and noise rejection compared to beamforming based methods. 
However, since they require a cross-spectrum matrix of sufficient, or sometimes full, rank, these methods fail for RIR data, which is considered to be a single snapshot or frame. 
Note that a cross-spectrum matrix constructed using multiple frames may also suffer from rank deficiency since reflections are correlated.
% FS - general and for SMAs. 
To overcome the problem of deficient rank cross-spectrum matrices, smoothing methods have been proposed as a preprocessing stage. 
The coherent signal subspace method \cite{wang1985coherent} and the array manifold interpolation method \cite{doron1994wavefield} are among the earlier published approaches used for frequency smoothing, i.e., a form of averaging along the frequency of the cross-spectrum matrix.
An efficient frequency-smoothing technique was later formulated for spherical microphone arrays \cite{khaykin2012acoustic}, which exploits a specific advantage of spherical array processing: 
the decoupling of the frequency and spatial components of sound field information, which enables frequency smoothing without the loss of spatial information.
While very useful, in addition to the loss of frequency resolution, frequency smoothing suffers from several limitations. 
One major limitation, for example, is that unless time separation is applied the total number of room reflections that can be localized using this method is limited by the number of microphones in the array. 
The application of a time-window as an initial step before applying frequency smoothing has been considered;
however, its performance is limited since short time windows lead to a loss of frequency resolution and limit the effect of frequency smoothing \cite{huleihel2013spherical}. 
% TS - SMAs 
Later, to overcome some of the shortcomings of frequency smoothing, time smoothing \cite{huleihel2013spherical} has been proposed. 
While time smoothing has been found to achieve better performance in terms of DOA estimation accuracy, it also suffers from some shortcomings. 
In addition to the inherent loss of temporal resolution, the method suffers from an amplification of noise at low frequencies since it includes, as an initial step, plane-wave decomposition in the frequency domain. 
This decomposition includes a division by radial functions that depend on the array type, which leads to ill-conditioning at low frequencies. 
The ill-conditioning may then lead to noise amplification and result in erroneous DOA estimations.
%new, addition of SLAs to previous work: 
Lately, multiple-input multiple-output (MIMO) systems, which employ a spherical loudspeaker array instead of a single loudspeaker, have been studied for room acoustics \cite{morgenstern2015theory,morgenstern2017design,morgenstern2017spatial}. 
Thanks to the additional spatial diversity, the direction of radiation (DOR) of reflections in a room (relative to the loudspeaker array center) can also be estimated, using the same methods used for DOA estimation. 

While the above methods were found to be very useful, they may all fail to separate reflections with equal, or nearly-equal, time delays. 
However, multiple reflections with nearly-equal time delays can arise in practice due to room reflections, and so their DOA estimation could be of potential interest. 
This paper proposes a new method for MIMO systems, referred to as `modal smoothing', in which smoothing over the channels, or spherical harmonic (SH) modes of the loudspeaker array, is performed with the aim of overcoming the shortcoming of previous methods, as detailed above.
%which could overcome this last shortcoming by performing smoothing over the channels of the spherical loudspeaker array. 
Preliminary results of this method were presented \cite{morgenstern2013enhanced}, showing that it also operates well when time separation of reflections is unavailable. 
This paper extends the results of Ref.~17, %\cite{morgenstern2013enhanced}, 
with the following additional contributions: 
\begin{itemize}
\item[(i)] A new simplified formulation of modal smoothing compared to its original formulation; % from Ref.~\cite{morgenstern2013enhanced}; 
the simplified formulation facilitates an analysis of the performance and limitations of modal smoothing based on characteristics of MIMO systems, which were presented in Ref.~14.%\cite{morgenstern2015theory}.} 
\item[(ii)] An extended simulation study; 
the study includes a configuration with room reflections with equal time delays, for which modal smoothing is shown to lead to accurate DOA estimations, while frequency smoothing is shown to fail. 
The study also presents the application of modal smoothing combined with frequency smoothing, which is shown to be beneficial for systems with loudspeaker arrays of low spatial resolution, or low SH order. 
\end{itemize}
The paper shows that employing modal smoothing can achieve successful decorrelation of room reflections with similar time delays, which, followed by the application of optimal processing methods such as MUSIC, could facilitate improved estimations of the DOAs of early reflections in a RIR. 

The paper is organized as follows. 
After a brief review of the MIMO system model and of frequency smoothing in Secs.~II and III, %\ref{sec:background} and \ref{sec:FS}, 
respectively, modal smoothing is derived for MIMO systems in Sec.~IV, and its performance and limitations are discussed.  %\ref{sec:MS}.  
The paper concludes with a simulation study in Sec.~V, %\ref{sec:simulations}, 
demonstrating the directional analysis provided by modal smoothing, frequency smoothing, and modal smoothing combined with frequency smoothing for an acoustic setup that includes room reflections with equal time delays.

\section{System model}\label{sec:background}
\justify % after each section and subsection

The model of a MIMO system that includes a spherical loudspeaker array and a spherical microphone array is presented in this section. 
The system is formulated for free field conditions, and is represented initially in the spatial domain and then in the SH domain. 
A model for a system positioned in a room is then formulated. 
Finally, the radial-function elimination is applied to the spherical microphone array, as it has been shown to be beneficial in room acoustics applications \cite{rafaely2004plane,morgenstern2015theory}. 
For further details regarding the complete formulation of MIMO systems, including a study of system properties and of limitations, the reader is referred to Refs.~14 and 15. 
%\cite{morgenstern2015theory} and \cite{morgenstern2017design}. 

\subsection{Space-domain model for free field}
\begin{figure}[H]
\centering
\includegraphics[clip, trim = {0 0 0 {5 mm}}, width = 1\linewidth]{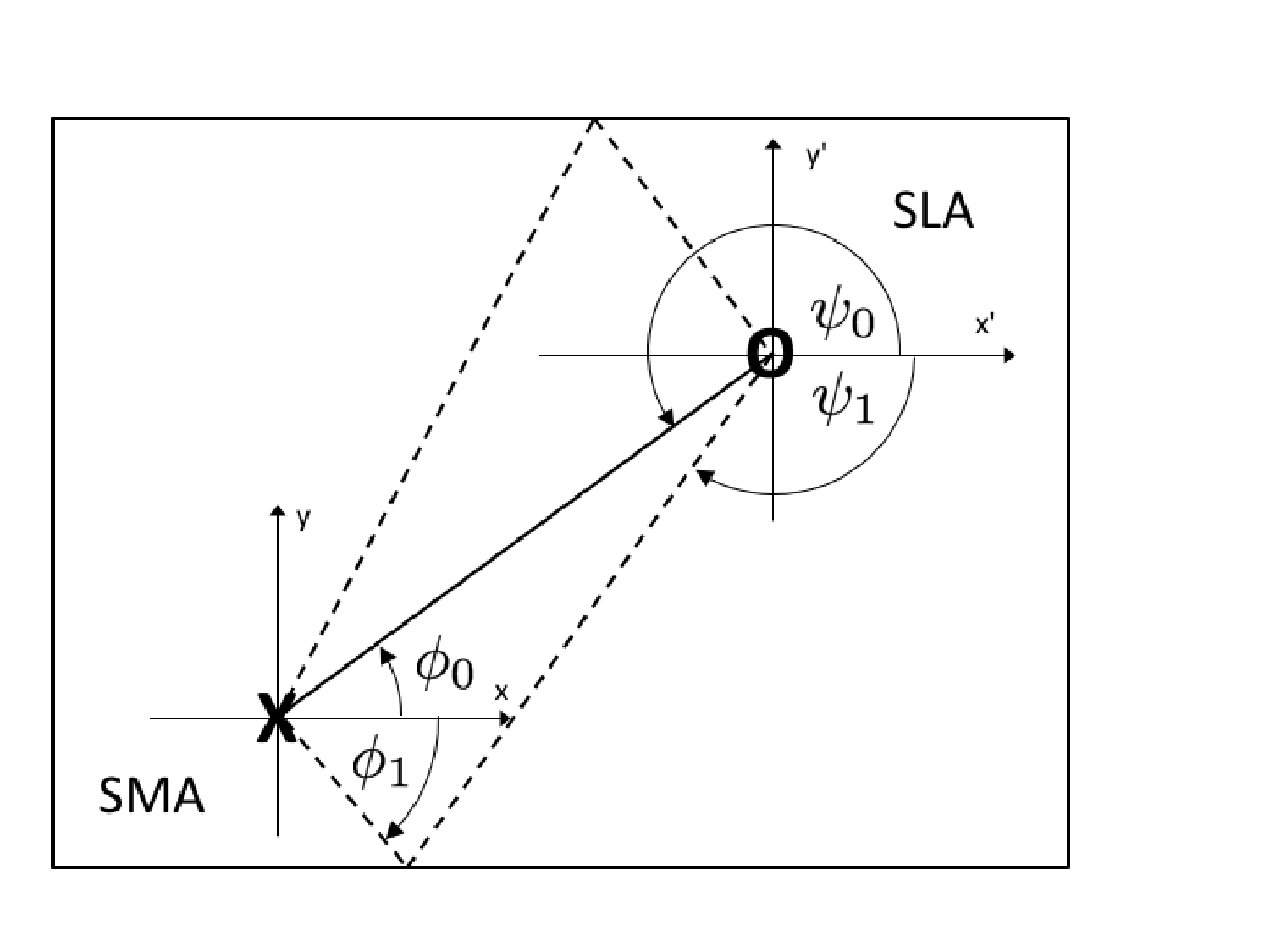}
\caption[System diagram in the x-y plane; O and X represent the spherical loudspeaker array and the spherical microphone array, respectively. The solid line represents direct sound and the dashed lines represent reflections (only two reflections are illustrated). $\psi_{0}$ and $ \phi_{0 }$ are the azimuth angles of the DOR and DOA for the direct sound, respectively. $\psi_{1}$ and $ \phi_{1}$ are the corresponding angles for the sound reflected by the wall at the bottom of the figure. In this case, $\beta_{0 } = \theta_{0 } = \beta_{1 } = \theta_{1 } =90^\circ$.]{System diagram in the x-y plane; O and X represent the spherical loudspeaker array and the spherical microphone array, respectively. The solid line represents direct sound and the dashed lines represent reflections (only two reflections are illustrated). $\psi_{0}$ and $ \phi_{0 }$ are the azimuth angles of the DOR and DOA for the direct sound, respectively. $\psi_{1}$ and $ \phi_{1}$ are the corresponding angles for the sound reflected by the wall at the bottom of the figure. In this case, $\beta_{0 } = \theta_{0 } = \beta_{1 } = \theta_{1 } =90^\circ$.}
\label{fig1}
\end{figure}

\justify
A system consisting of a spherical loudspeaker array with radius $r_L$ and $S$ loudspeakers and a spherical microphone array with radius $r_M$ and $R$ microphones is considered. 
The MIMO system is represented in the frequency domain, for which modal smoothing will also be derived in the next section. 
A time-domain formulation is also possible, similar to that found in Ref.~13. %\cite{huleihel2013spherical}. 
Due to its adjustable directivity, a spherical loudspeaker array can produce arbitrary sound fields. % around a SMA. 
This is achieved by weighting the loudspeaker array input signal, $s(\omega)$, where $\omega$ is the angular frequency, with complex beamforming coefficients before applying it to the loudspeakers. 
In this case, the pressure at the microphone array can be written as \cite{morgenstern2017design}:
\begin{eqnarray}\label{eq:bck1s}
\tilde{\bm p}(\omega) & = &  \tilde{\bm H}(\omega) \tilde{\bm \gamma}(\omega) s(\omega), 
\end{eqnarray}
where 
\begin{eqnarray}\label{eq:bck2s}
\tilde{\bm p}(\omega) &=&  [p_{1}(\omega), \,p_{2}(\omega),\,p_{3}(\omega),\, ...,\, p_{R}(\omega)]^\mathrm{T} 
\end{eqnarray}
is a $R \times 1$ vector that holds the pressure of the microphones, and $(\cdot)^\mathrm{T}$ is the transpose operator.  
$\tilde{\bm H}(\omega)$ is the $R\times S$ system transfer matrix whose elements hold the transfer functions for all loudspeaker and microphone combinations, and 
\begin{eqnarray}\label{eq:bck3s}
\tilde{\bm \gamma}(\omega)& = &[\gamma_{1}(\omega), \,\gamma_{2}(\omega),\,\gamma_{3}(\omega),\, ...,\, \gamma_{S}(\omega)]^\mathrm{T}
\end{eqnarray}
is a $S\times 1$ vector that holds the loudspeaker array beamforming coefficients.
Note that Eq.~\eqref{eq:bck1s} assumes a single-frequency sound field in steady state, which could be extended to a broadband model by adding contributions from multiple frequencies. 
Therefore, $\tilde{\bm p}(\omega)$ from Eq.~\eqref{eq:bck1s} represents the complex amplitude of the sound pressure at the microphones.
For a block diagram of a MIMO system, including beamforming, the reader is referred to Ref.~15. %\cite{morgenstern2017design}.
In free field, and assuming that the distance between the centers of the arrays is large compared to their radii, the pressure radiated by the loudspeaker array can be approximated by a planar wavefront in the region of space occupied by the microphone array \cite{morgenstern2014farfield}. 
In this case, $\tilde{\bm H}(\omega)$ can be written as: 
\begin{eqnarray}\label{eq:bck4s}
\tilde{ \bm H}(\omega) & = &  \bm Y_M \bm B(\omega) \bm y^\mathrm{H}(\bm \theta_0) \bm y(\bm \beta_0) \bm G(\omega)  \bm Y_L^\mathrm{H}  \lambda_0(\omega) , 
\end{eqnarray}
where $\lambda_0(\omega)= e^{\frac{i\omega r_0}{c}}/r_0$, with $r_0$ being the distance between the centers of the arrays and $c$ being the speed of sound, accounts for the phase and the attenuation due to the propagation from the loudspeaker array center to the microphone array center. 
$(\cdot)^\mathrm{H}$ is the conjugate transpose operator, and $\bm Y_L$ is an $S\times (N_L+1)^2$ matrix given by: 
\begin{eqnarray}\label{eq:bck4s2}
\bm Y_L & = & [\bm y^\mathrm{H}(\bm \beta^L_1), \, \bm y^\mathrm{H}(\bm \beta^L_2), \,... , \, \bm y^\mathrm{H}(\bm \beta^L_S)], 
\end{eqnarray}
with 
\begin{eqnarray}\label{eq:bck8s}
\bm y(\bm \beta) & = & [Y_0^0(\bm \beta), \,Y_1^{-1}(\bm \beta),\,Y_1^0(\bm \beta),\,Y_1^1(\bm \beta),\, ... ,\, Y_{N_L}^{N_L}(\bm \beta)]
\end{eqnarray}
being a $1\times (N_L+1)^2$ vector that holds SH functions, $Y_n^m(\bm \beta)$ \cite{driscoll1994computing}, evaluated at elevation angle $\beta$ and azimuth angle $\psi$ with respect to the loudspeaker array center. 
$\bm \beta^L_1, \,\bm \beta^L_2 ,\,... \, \bm \beta^L_S$ are the spatial angles that point at the loudspeakers with respect to the array center. 
$N_L$ is referred to as the loudspeaker array SH order, and should tend to infinity for an exact representation of the sound field. 
However, a finite order is chosen in practice, as will be discussed in the following section.
$\bm G(\omega)$ is a $(N_L+1)^2\times (N_L+1)^2$ diagonal matrix given by: 
\begin{eqnarray}\label{eq:bck5}
\bm G(\omega) & = & \text{diag}[g_0\left( \frac{\omega r_L}{c}   \right)  ,\,g_1\left( \frac{\omega r_L}{c}   \right),\,g_1\left( \frac{\omega r_L}{c}   \right),\,g_1\left( \frac{\omega r_L}{c}   \right),\, ... ,\,g_{N_L}\left( \frac{\omega r_L}{c}   \right)],   
\end{eqnarray}
which holds coefficients that correspond to the spherical loudspeaker array type and parameters. 
In particular, the loudspeaker membranes are modeled as spherical caps \cite{williams1999fourier}. 
$g_n\left( \frac{\omega r_L}{c}   \right)$ are given by: 
\begin{eqnarray}\label{eq:bck6}
g_{n}\left( \frac{\omega r_L}{c}   \right) & = &   \rho c r^2_L (-i)^{n+1} \left( j_{n}\left( \frac{\omega r_L}{c}   \right) - \frac{j'_{n}\left( \frac{\omega r_L}{c}   \right)}{h'_{n}\left( \frac{\omega r_L}{c}   \right)}h_{n}\left( \frac{\omega r_L}{c}   \right) \right) q_n.   
\end{eqnarray}
$\rho$ is the air density, 
$i^2=-1$,  
$j_{n}(\cdot)$ and $j'_{n}(\cdot)$ are the spherical Bessel function of order $n$ and its derivative, respectively,
$h_{n}(\cdot)$ and $h'_{n}(\cdot)$ are the spherical Hankel function of the first kind of order $n$ and its derivative, respectively, 
and 
\begin{eqnarray}\label{eq:bck7}
q_n  = 
\begin{cases}
4\pi^2(1-\cos\alpha), \, \, n =0, \\
\frac{4\pi^2}{2n+1}   \begin{pmatrix} P_{n-1}(\cos\alpha)- P_{n+1}(\cos\alpha) \end{pmatrix}, \, \, n >0,
\end{cases}
\end{eqnarray}
with $P_{n}(\cdot)$ being the Legendre polynomial of order $n$, and $\alpha$ being the aperture angle that corresponds to the size of the loudspeaker units relative to the array radius. 
$\bm G(\omega)\bm Y_L^{\mathrm{H}}$ represents the radiation from the loudspeaker membranes.
Note that $g_{n}\left( \frac{\omega r_L}{c}   \right) $ from Eq.~\eqref{eq:bck5} is formulated for a signal $s(\omega)$ that represents voltage. 
An additional $\frac{\omega}{c}$ term should be added to the last equation if $s(\omega)$ represents loudspeaker membrane velocity, instead of voltage \cite{rafaely2011optimal}. 
Note also that $g_{n}\left( \frac{\omega r_L}{c}   \right)$ is defined in accordance with the steady state solution from Ref.~19, %\cite{williams1999fourier}, 
employing a Fourier basis with negative frequency \cite{tourbabin2015consistent}.
The rest of the equations in this paper are also in accordance with this convention.
$\bm y(\bm \beta_0)$ is the loudspeaker array steering vector, defined as in Eq.~\eqref{eq:bck8s}, but using $\bm \beta_0 $, which is the DOR.  
The loudspeaker array steering vector may be more complex in practice.
However, $\bm y(\bm \beta_0)$ can approximately represent this vector under some simplifying assumptions, as discussed in the following section.
In a similar manner, $\bm y(\bm \theta_0)$ is a $1\times (N_M+1)^2$ steering vector given as in Eq.~\eqref{eq:bck8s}, but using $N_M$, the microphone array SH order, and $\bm \theta_0$, the corresponding DOA written using the vector notation of $\bm \theta_0 = (\theta_0, \phi_0)$, with elevation angle $\theta_0$ and azimuth angle $\phi_0$ with respect to the microphone array center. 
Note that, as in the case of the spherical loudspeaker array, $N_M$ should tend to infinity for an exact representation of the sound field, but a finite order is chosen in practice, as will be explained in the following section. 
$\bm B(\omega)$ is a $(N_M+1)^2\times (N_M+1)^2$ diagonal matrix given by: 
\begin{eqnarray}\label{eq:bck9}
\bm B(\omega) &=& \text{diag}[b_0\left( \frac{\omega r_M}{c}   \right),\,b_1\left( \frac{\omega r_M}{c}   \right),\,b_1\left( \frac{\omega r_M}{c}   \right),\,b_1\left( \frac{\omega r_M}{c}   \right),\,...  ,\,b_{N_M}\left( \frac{\omega r_M}{c}   \right)],    
\end{eqnarray}
which holds coefficients that correspond to the spherical microphone array type and parameters given by \cite{rafaely2004plane}:
\begin{eqnarray}\label{eq:bck10}
b_{n}\left( \frac{\omega r_M}{c}   \right) & = & 4\pi(-i)^{n}  \left( j_{n}\left( \frac{\omega r_M}{c}   \right) - \frac{j'_{n}\left( \frac{\omega r_M}{c}   \right)}{h'_{n}\left( \frac{\omega r_M}{c}   \right)}h_{n}\left( \frac{\omega r_M}{c}   \right) \right). 
\end{eqnarray}
Finally, $\bm Y_M$ is an $R\times (N_M+1)^2$ matrix defined as in Eq.~\eqref{eq:bck8s}, but for $N_M$ and for the spatial angles that point at the microphones with respect to the array center, $\bm\theta^M_1, \,\bm\theta^M_2, \,...,\,\bm \theta^M_R$.    
Fig.~\ref{fig1} shows a schematic illustration of the system. 
Additional information regarding the system presented in the figure will be provided in Sec.~II B. 
% former place of Fig. 1

\subsection{Spherical harmonic model for free field}
\justify

An alternative representation of the system in the SH domain is now given, since it is simpler and more appropriate for spatial analysis with spherical arrays \cite{morgenstern2015theory}. 
Transfer functions are formulated between the $(N_L+1)^2$ SH channels of the loudspeaker array and the $(N_M+1)^2$ channels of the microphone array, instead of between the $S$ loudspeakers and the $R$ microphones.  
At this stage, the loudspeaker and microphone SH orders are assumed to maintain $(N_L+1)^2\leq S$ and $(N_M+1)^2\leq R$, and the elements in each array are assumed to be distributed according to known sampling schemes, such as nearly-uniform or Gaussian distributions \cite{rafaely2005analysis}. 
As discussed earlier, an infinite-order sound field is typically produced by the spherical loudspeaker array, and, similarly, an infinite-order sound field is expected around the microphone array.
However, the effective SH orders of the sound fields are limited in practice when measured on a spherical surface around the arrays. 
The effective SH orders are determined by the attenuation of high sound field SH orders due to the radial functions $g_n\left( \frac{\omega r_L}{c}   \right) $ and $b_n\left( \frac{\omega r_M}{c}   \right)$ for the loudspeaker and microphone arrays, respectively.   
In particular, this attenuation becomes significant for $n>> \frac{\omega r_L}{c}  $ for the loudspeaker array and $n>> \frac{\omega r_M}{c}$ for the microphone array \cite{rafaely2009spherical}. 
Therefore, aliasing error can be kept to a minimum if the effective SH orders are lower than $N_L$ and $N_M$ for the loudspeaker and microphone arrays, respectively. 
With no aliasing and with the sound fields assumed to be of finite SH orders, the effect of the physical construction of the arrays can be removed, and $\bm y(\bm \beta)$ and $\bm y(\bm \theta)$ can correctly represent the loudspeaker and microphone array steering vectors in the SH domain, respectively.  
For a full formulation and study of spatial aliasing in MIMO systems with spherical arrays, the reader is referred to Ref.~15. %\cite{morgenstern2017design}.}  
When the system is represented in the SH domain, the pressure at the microphone array can be written as \cite{morgenstern2017design}:
\begin{eqnarray}\label{eq:bck1}
\bm p(\omega) & = &  \bm H(\omega) \bm \gamma(\omega) s(\omega), 
\end{eqnarray}
where 
\begin{eqnarray}\label{eq:bck2}
\bm p(\omega) &=&  [p_{00}(\omega), \,p_{1(-1)}(\omega),\,p_{10}(\omega),\, p_{11}(\omega),\, ...,\, p_{N_MN_M}(\omega)]
\end{eqnarray}
is a $(N_M+1)^2\times 1$ vector that holds the pressure of the microphone array SH channels, whose elements are described by two indices, e.g., $p_{nm}(\omega)$, where $n$ and $m$ are the SH order and the SH degree, respectively.  
\begin{eqnarray}\label{eq:bck3}
\bm \gamma(\omega)& = &[\gamma_{00}(\omega), \,\gamma_{1(-1)}(\omega),\,\gamma_{10}(\omega),\, \gamma_{11}(\omega),\, ...,\, \gamma_{N_LN_L}(\omega)]^\mathrm{T}
\end{eqnarray}
is a $(N_L+1)^2$ vector that holds the loudspeaker array beamforming coefficients in the SH domain, and 
 \begin{eqnarray}\label{eq:bck4}
 \bm H(\omega) & = &  \bm B(\omega) \bm y^\mathrm{H}(\bm \theta_0) \bm y(\bm \beta_0) \bm G(\omega)   \lambda_0(\omega).  
\end{eqnarray}
For a discussion of the properties of $\bm H(\omega)$ and its relation to the space domain system $\tilde{ \bm H}(\omega)$, the reader is referred to Ref.~14. %\cite{morgenstern2015theory}. 

\subsection{Model for MIMO system in a room}
\justify

A MIMO system positioned in a room can be represented as a summation of multiple free-field systems for the individual room reflections \cite{allen1979image}. 
In this case, $\bm H(\omega)$ from Eq.~\eqref{eq:bck5} can be written as \cite{morgenstern2015theory}: 
\begin{eqnarray}\label{eq:bck11}
\bm H(\omega) & = & \sum_{l= 1}^{L} \bm B(\omega) \bm y^\mathrm{H}(\bm \theta_l) \bm y(\bm \beta_l)\bm G(\omega) \lambda_l (\omega)  \\
 & = & \bm B(\omega) \bm Y^\mathrm{H}(\bm \Theta) \bm \Lambda(\omega) \bm Y(\bm \Phi) \bm G(\omega),   \label{eq:bck12}
\end{eqnarray}
where $\bm y(\bm \beta_l)$ and $\bm y(\bm \theta_l)$ are the loudspeaker and microphone array steering vectors of reflection $l$, respectively, with the corresponding DOR, $\bm \beta_l$, and DOA, $\bm \theta_l$, $\lambda_l(\omega)$ accounts for the phase and attenuation of that reflection due to propagation and absorption by the walls, and $L$ is the number of significant room reflections.  
In particular, note that the DORs of each reflection, or image source, take into account both the displacement and the mirroring of that source. 
The alternative matrix representation of the system in Eq.~\eqref{eq:bck12} is given using the $L \times (N_L+1)^2$ loudspeaker array steering matrix:
\begin{eqnarray}\label{eq:bck13}
\bm Y(\bm \Phi) & = & [\bm y(\bm \beta_1)^\mathrm{H},\, \bm y(\bm \beta_2)^\mathrm{H}, ..., \bm y(\bm \beta_L)^\mathrm{H}]^\mathrm{H}, 
\end{eqnarray}
the $L \times (N_M+1)^2$ microphone array steering matrix:
\begin{eqnarray}\label{eq:bck14}
\bm Y(\bm \Theta) & = & [\bm y(\bm \theta_1)^\mathrm{H},\, \bm y(\bm \theta_2)^\mathrm{H},\, ...,\, \bm y(\bm \theta_L)^\mathrm{H}]^\mathrm{H}, 
\end{eqnarray}
and a $L \times L$ diagonal matrix:
\begin{eqnarray}\label{eq:bck15}
\bm \Lambda(\omega)  & = & \text{diag}[\lambda_1(\omega) ,\, \lambda_2(\omega) , \,..., \,\lambda_L(\omega) ],
\end{eqnarray}
which holds $\lambda_l(\omega)$ for all reflections. 
An important characteristic of the room system transfer matrix is that its rank is bounded by the number of significant reflections or by its  dimensions (the lower of the two)  \cite{morgenstern2015theory}, i.e., $\text{min}(L, (N_L + 1)^2, (N_M + 1)^2)$, which will be useful in the formulation of modal smoothing in the next section. 
Note that the number of reflections in a room can be high; 
thus, the limitation on matrix rank presented in this section constrains the method to work in practice only when the reflections are relatively sparse, i.e.,~for the early reflections. 
This limitation will be discussed in more detail in Sec.~IV.
The reader is referred again to Fig.~\ref{fig1} for an illustration of the direct sound and some room reflections.
For ease of illustration, the system diagram is given in the $x-y$ plane and both of the arrays are positioned at the same height. 
The coordinate systems of the arrays are aligned with the walls of the room and, therefore, the DOA and DOR elevation angles are both set at $\theta_0 = \beta_0 = 90^\circ$. In the figure, the solid line represents the direct sound, and $\phi_0$ and $\psi_0$ are the corresponding azimuth angles of the DOA and DOR, respectively. 
The dashed lines represent two wall reflections, with the corresponding azimuth angles, $\phi_0$ and $\psi_0$, of one of these reflections.

\subsection{Radial-function elimination}
\justify

As a last step in the formulation of the system model, plane-wave decomposition \cite{rafaely2004plane}, or radial-function elimination \cite{morgenstern2017design}, is applied to the microphone array, which facilitates a representation of the sound field using plane waves. 
In particular, applying plane-wave decomposition to spherical microphone arrays facilitates the decoupling of the frequency and angular components of $\bm p(\omega)$, which is required for both frequency smoothing and time smoothing. 
While it is not required for modal smoothing, it is nevertheless applied, since it leads to improved performance in the application presented in this paper.
On the other hand, radial-function elimination is not applied to the loudspeaker array for this specific application, although it has been reported to lead to improved performance in other room acoustics applications \cite{morgenstern2015theory, morgenstern2017spatial}.  
A system transfer matrix that includes plane-wave decomposition at the microphone array, $\bm A(\omega)$, is defined as $\bm H(\omega)$ from Eq.~\eqref{eq:bck12}, but without $\bm B(\omega)$, i.e., 
\begin{eqnarray}\label{eq:bck16}
\bm A(\omega) & = & \sum_{l= 1}^{L} \bm y^\mathrm{H}(\bm \theta_l) \bm y(\bm \beta_l)\bm G(\omega) \lambda_l(\omega) \\
 & = &  \bm Y^\mathrm{H}(\bm \Theta) \bm \Lambda(\omega) \bm Y(\bm \Phi) \bm G(\omega).  
\end{eqnarray}
The plane-wave density at the microphone array SH channels, written in vector notation as $\bm a(\omega)$, is now calculated as $\bm p(\omega)$ from Eq.~\eqref{eq:bck1}, but using $\bm A(\omega)$ instead of $\bm H(\omega)$, i.e., as:
\begin{eqnarray}\label{eq:bck17}
\bm a(\omega) & = &  \bm A(\omega) \bm \gamma(\omega) s(\omega). 
\end{eqnarray}
Note that the estimation of $\bm A(\omega)$ involves the inversion of $\bm B(\omega)$, and therefore introduces ill conditioning at low frequencies, which imposes limitations on the practical operating frequency range of the system \cite{rafaely2005analysis,morgenstern2017design}. 
This ill conditioning may amplify errors in the system transfer matrix, namely, system identification errors. 
In fact, it should be mentioned that $\bm p(\omega)$ and $\bm a(\omega)$ from Eqs.~\eqref{eq:bck1} and \eqref{eq:bck17}, respectively, are not complete, as they lack a term that corresponds to system identification errors, which was omitted for simplicity.  
For a MIMO system model, a generalized additive error model is employed to address such errors, as in Refs.~14, 15, and 16, %\cite{morgenstern2015theory, morgenstern2017spatial, morgenstern2017design}, 
since it facilitates an analysis that is not constrained to specific system identification methods. 
An error matrix, $\bm N(\omega)$, is added to the the system transfer matrix before applying beamforming, i.e., $\bm H(\omega) + \bm N(\omega)$ is substituted in Eq.~\eqref{eq:bck1} instead of $\bm H(\omega)$. 
Similarly, $\bm A(\omega) + \bm B^{-1}(\omega) \bm N(\omega)$ should be substituted into Eq.~\eqref{eq:bck17}, instead of $\bm A(\omega)$. 
The generation of such errors will be discussed in more detail in Sec.~V. %\ref{sec:simulations}.
Finally, even though system identification errors are assumed throughout the rest of the paper, they are omitted from all equations henceforward, since the focus of both the previous and the modal smoothing methods is on restoring the rank of the signal cross-spectrum matrix, as will be demonstrated in the next sections.  

\section{Frequency smoothing}\label{sec:FS}
\justify 

Frequency smoothing is designed for spherical microphone arrays as a solution to the problem of the inherent unit rank of the cross-spectrum matrix based on impulse response data. 
It is typically done as an initial step before employing optimal array processing methods such as MUSIC \cite{khaykin2012acoustic}. 
This section briefly presents frequency smoothing, as it will serve as an example of previous methods and then be compared to modal smoothing in Sec.~V. %\ref{sec:simulations}.

The original formulation of frequency smoothing assumes that the sound field surrounding the microphone array is composed of $L$ plane waves \cite{khaykin2012acoustic}. 
As mentioned earlier, $L$ can be very high in practice. In this case, time windowing could be applied to limit the number of reflections in the RIR. This is discussed in more detail in Sec.~IV.
To relate between the model employed in Ref.~4 and the system model presented in the previous section, the following $L\times 1$ signal vector, $\bm s(\omega)$, is defined for a selected loudspeaker array beamforming vector $\bm \gamma(\omega)$. 
$\bm s(\omega)$ is defined as:
\begin{eqnarray}\label{eq:FS1}
\bm s(\omega) & = &  \bm \Lambda(\omega) \bm Y(\bm \Phi)  \bm G(\omega) \bm \gamma(\omega) s(\omega).  
\end{eqnarray}
In particular, $s(\omega)$ is assumed to be unity, and is omitted from the rest of the equations henceforward, since the methods are developed for RIR data. 
$\bm a(\omega)$ from Eq.~\eqref{eq:bck17} can now be written as in Ref.~4:%\cite{khaykin2012acoustic}: 
\begin{eqnarray}\label{eq:FS2}
\bm a(\omega) & = & \bm Y^\mathrm{H}(\bm \Theta) \bm s(\omega).   
\end{eqnarray}
Using this notation of $\bm a(\omega)$, the modal cross-spectrum matrix, $\bm S_{\bm a}(\omega)$, should be calculated as: 
\begin{eqnarray}\label{eq:FS3}
\bm S_{\bm a}(\omega) & = & E[ \bm a(\omega)  \bm a^\mathrm{H} (\omega) ] \nonumber \\
& = &  \bm Y^\mathrm{H}(\bm \Theta) \bm S_{\bm s}(\omega) \bm Y(\bm \Theta),
\end{eqnarray}
where $E[\cdot]$ denotes the statistical expectation, and $\bm S_{\bm s}(\omega)$ is the signal cross-spectrum matrix. 
Since RIR data is used, only one realization of $\bm a(\omega)$ is available, so that the expectation cannot really be computed.  
Instead, $\bm S_{\bm a}(\omega)$ is estimated using $ \bm a(\omega)  \bm a^\mathrm{H} (\omega)$, which results in an inherent unit rank. 
To cope with this rank deficiency, the modal cross-spectrum matrix is smoothed to produce $\tilde{\bm S}_{\bm a} $, by averaging over $Q$ frequency bands. 
$\tilde{\bm S}_{\bm a} $ is given by: 
\begin{eqnarray}\label{eq:FS4}
\tilde{\bm S}_{\bm a} & = & \frac{1}{Q} \sum_{q = 1}^{Q} \bm a(\omega_q)  \bm a^\mathrm{H} (\omega_q) \nonumber \\
& = & \bm Y^\mathrm{H}(\bm \Theta) \tilde{\bm S}_{\bm s} \bm Y(\bm \Theta),
\end{eqnarray} 
where $\tilde{\bm S}_{\bm s}$ is the smoothed signal across-spectrum matrix. 
It is evident that at least $Q\geq L$ frequency bands are required for correctly estimating $L$ DOAs, and that the microphone array steering matrix remains unchanged after smoothing, as desired. 
MUSIC can now be employed using the smoothed modal cross-spectrum matrix for estimating the DOAs of the $L$ plane waves, as described in Ref.~4, %\cite{khaykin2012acoustic}, 
and as will be discussed in Sec.~V. %\ref{sec:simulations}. 
In practice, a much higher number of frequency bands is used for computing $\tilde{\bm S}_{\bm s}$ \cite{khaykin2012acoustic}. 
When a high number of frequency bands is used, the number of room reflections may increase, because short time windows can no longer be applied. 
In particular, if the number of plane waves is higher than the dimension of the cross-spectrum matrix from Eq.~\eqref{eq:FS4}, which is determined by the SH order of the microphone array, frequency smoothing is expected to fail. 
This is a major downside of frequency smoothing.

While frequency smoothing is useful, it imposes an inevitable loss of frequency resolution, and may lead to failure to distinguish between reflections with equal time delays. 
This will be demonstrated using simulated RIR data in Sec.~V. %\ref{sec:simulations}. 
This can also be seen analytically, by studying the structure of the signal vector. 
For example, consider a simple configuration with $L = 2$ reflections with different DOAs, but equal time delays. 
In this case, $\bm s(\omega)$ can be assumed to be spanned by a single vector for all frequencies, since $\lambda_l(\omega)$ from Eq.~\eqref{eq:bck11} is the same for the two reflections, which implies that $\tilde{\bm S}_{\bm s}$ will have unit rank in this case. 
Similar behavior is expected for reflections with nearly-equal time delays; 
the effective rank \cite{roy2007effective} of $\tilde{\bm S}_{\bm s}$ may be close to one, and so it may not be feasible to distinguish between the two reflections using MUSIC in this case, considering that a noise component is also added to the cross-spectrum matrix. 
Similarly, time smoothing is also expected to fail in the case of reflections with equal time delays. 

\section{Modal smoothing}\label{sec:MS}
\justify % after each section and subsection

In this section, modal smoothing for MIMO systems is developed in order to overcome the rank deficiency problem of the smoothed modal cross-spectrum that arises in the previous methods when there are different reflections with the same time delay. 

Multiple cross-spectrum matrices can be generated at a single frequency in the case of a MIMO system, due to adjustable directivity of the spherical loudspeaker array, as demonstrated in Eq.~\eqref{eq:FS1}. 
This is exploited in the formulation of modal smoothing, in which signal vectors (as in Eq.~\eqref{eq:FS1}) and corresponding modal vectors (as in Eq.~\eqref{eq:FS2}) are defined for each of the loudspeaker array SH channels. 
For example, an omnidirectional source can be modeled by choosing $\bm \gamma(\omega) = \bm d_{00} = [1, 0, 0, ..., 0]^\mathrm{T}$ to define $\bm s_{00}(\omega)$ from Eq.~\eqref{eq:FS1}, which then gives a corresponding modal vector, $\bm a_{00}(\omega)$, when substituted in Eq.~\eqref{eq:FS2}. 
Modal vectors of other loudspeaker array SH channels are calculated similarly, by choosing $\bm \gamma(\omega) = \bm d_{nm}$, where $\bm d_{nm}$ is defined similarly to $\bm d_{00}$, but with zeros everywhere except for the element that corresponds to loudspeaker array SH channel of order $n$ and degree $m$. 
A smoothed modal cross-spectrum matrix, $\tilde{\bm S}_{\bm A}(\omega)$, can now be constructed using the $(N_L+1)^2$ modal vectors. 
$\tilde{\bm S}_{\bm A}(\omega)$ is defined as:
\begin{eqnarray}\label{eq:MS1}
\tilde{\bm S}_{\bm A}(\omega) & = & \frac{1}{(N_L+1)^2} \sum_{n = 0}^{N_L} \sum_{m = -n}^n \bm a_{nm}(\omega) \bm a^{\mathrm{H}}_{nm}(\omega)
\nonumber \\
& = &  \bm Y^\mathrm{H}(\bm \Theta)  \tilde{\bm S}_{\bm S}(\omega)   \bm Y(\bm \Theta), 
\end{eqnarray}
where $\tilde{\bm S}_{\bm S}(\omega)$ is the smoothed signal cross-spectrum matrix given by: 
\begin{eqnarray}\label{eq:MS2}
\tilde{\bm S}_{\bm S}(\omega)  & = &\frac{1}{(N_L+1)^2} \sum_{n = 0}^{N_L} \sum_{m = -n}^n 
 \bm \Lambda(\omega) \bm Y(\bm \Phi)  \bm G(\omega) 
 \bm d_{nm}  \bm d_{nm}^\mathrm{H}  \bm G^\mathrm{H}(\omega) \bm Y^\mathrm{H}(\bm \Phi) \bm \Lambda^\mathrm{H}(\omega) \nonumber \\
& = & \frac{ 1}{(N_L+1)^2}\bm  \Lambda(\omega) \bm Y(\bm \Phi) \bm G(\omega)
\left( \sum_{n = 0}^{N_L} \sum_{m = -n}^n \bm d_{nm} 
\bm d_{nm}^\mathrm{H}\right)  
\bm G^\mathrm{H}(\omega) \bm Y^\mathrm{H}(\bm \Phi) \bm \Lambda^\mathrm{H}(\omega) \\
& = & \frac{ 1}{(N_L+1)^2} \bm \Lambda(\omega) \bm Y(\bm \Phi) \bm G(\omega)
\bm G^\mathrm{H}(\omega)  \bm Y^\mathrm{H}(\bm \Phi) \bm \Lambda^\mathrm{H}(\omega), \label{eq:MS3}
\end{eqnarray} 
with $(\cdot)^*$ denoting the complex conjugate. 
In particular, the transition from Eq.~\eqref{eq:MS2} to Eq.~\eqref{eq:MS3} is done using the relation 
\begin{eqnarray}\label{eq:MS4}
\sum_{n = 0}^{N_L} \sum_{m = -n}^n \bm d_{nm} \bm d_{nm}^\mathrm{H}  =&  \bm I, 
\end{eqnarray}
where $\bm I$ is the $(N_L+1)^2 \times (N_L+1)^2$ identity matrix. 
As can be seen in Eq.~\eqref{eq:MS1}, the microphone array steering matrix does not change as a result of smoothing in the smoothed cross-spectrum matrix, because the loudspeaker array directivity does not affect this matrix, as seen in Eq.~\eqref{eq:FS2}. 
This is a key element in modal smoothing, and is facilitated by the use of a spherical loudspeaker array; 
the SH channels correspond to modal sources that are concentric, i.e., sources that share a common center - the loudspeaker array center. 
This is a significant advantage of modal smoothing compared to other forms of spatial smoothing, such as smoothing along the (typically nonconcentric) loudspeakers of an array.  
This structure of the microphone array steering matrix will also be preserved for a time domain formulation \cite{morgenstern2013enhanced}.
Furthermore, note that using Eq.~\eqref{eq:MS3}, $\tilde{\bm S}_{\bm A}(\omega)$ from Eq.~\eqref{eq:MS1} can be written as: 
\begin{eqnarray}\label{eq:MS5}
\tilde{\bm S}_{\bm A}(\omega) & = & \frac{1}{(N_L+1)^2} \bm A(\omega) \bm A^{\mathrm{H}}(\omega). 
\end{eqnarray}
This is a very useful result. 
$\bm A(\omega)$, the MIMO system transfer matrix, has been studied in Ref.~14, %\cite{morgenstern2015theory}, 
providing insight into the expected performance and limitations of modal smoothing: 
its rank was proposed as an estimator for the number of significant reflections, and an analysis of this matrix's singular values was provided. 
The singular values were shown to be comprised of two distinct groups: the signal singular values, equal in number to the number of reflections (when using high enough SH orders of the arrays), and the noise singular values, which include all singular values that are not included in the first group. 
It has been shown that there is a distinct separation between the signal singular values and the noise singular values, which could facilitate the estimation of the number of reflections based on the rank of $\bm A(\omega)$. 
Nevertheless, limitations on the performance are imposed by the configuration parameters (such as the SH orders of the arrays, the frequencies under analysis, and the power of the errors). 
MUSIC operates similarly, exploiting this same separation between the signal singular values and the noise singular values for obtaining the signal and noise subspaces. 
Especially, and following the relation between $\tilde{\bm S}_{\bm A}(\omega)$ and $\bm A(\omega)$ as in Eq.~\eqref{eq:MS5}, the eigenvalues of $\tilde{\bm S}_{\bm A}(\omega)$ are proportional to the absolute square of the singular values of $\bm A(\omega)$. 
Thus, signal eigenvalues and noise eigenvalues can be defined for $\tilde{\bm S}_{\bm A}(\omega)$, corresponding to the signal singular values and the noise singular values of $\bm A(\omega)$, respectively. 
Expanding on this relation between the eigenvalues of $\tilde{\bm S}_{\bm A}(\omega)$ and singular values of $\bm A(\omega)$, the performance of modal smoothing, i.e., the ability to successfully restore the rank of $\tilde{\bm S}_{\bm A}(\omega)$, is expected to be similar to that of the estimation of the number of reflections, which was extensively studied in Ref.~14. %\cite{morgenstern2015theory}. 
In particular, one limitation that arises from the analysis in Ref.~14 %\cite{morgenstern2015theory} 
is that $N_L$ should be chosen such that $(N_L+1)^2\geq L$ for successful decorrelation of $L$ reflections.  
As mentioned earlier, the number of reflections can be very high in practice, especially if the number of room modes is high. 
In this case, time windowing could be applied as a preprocessing stage to limit the number of reflections. 
An example that includes time windowing is presented in the next section.

\section{Simulation studies}\label{sec:simulations}
\justify % after each section and subsection

In this section, the proposed method is applied on simulated RIRs to illustrate the advantage of modal smoothing over previous smoothing methods in the case of multiple reflections with equal time delays. 
The option of applying modal smoothing combined with frequency smoothing is also proposed for spherical loudspeaker arrays of low SH orders. 

\subsection{Setup}% setup
\justify 

\begin{figure}[H]
\centering
\includegraphics[clip, trim = {0 0 0 {0 mm}}, width = 1\linewidth]{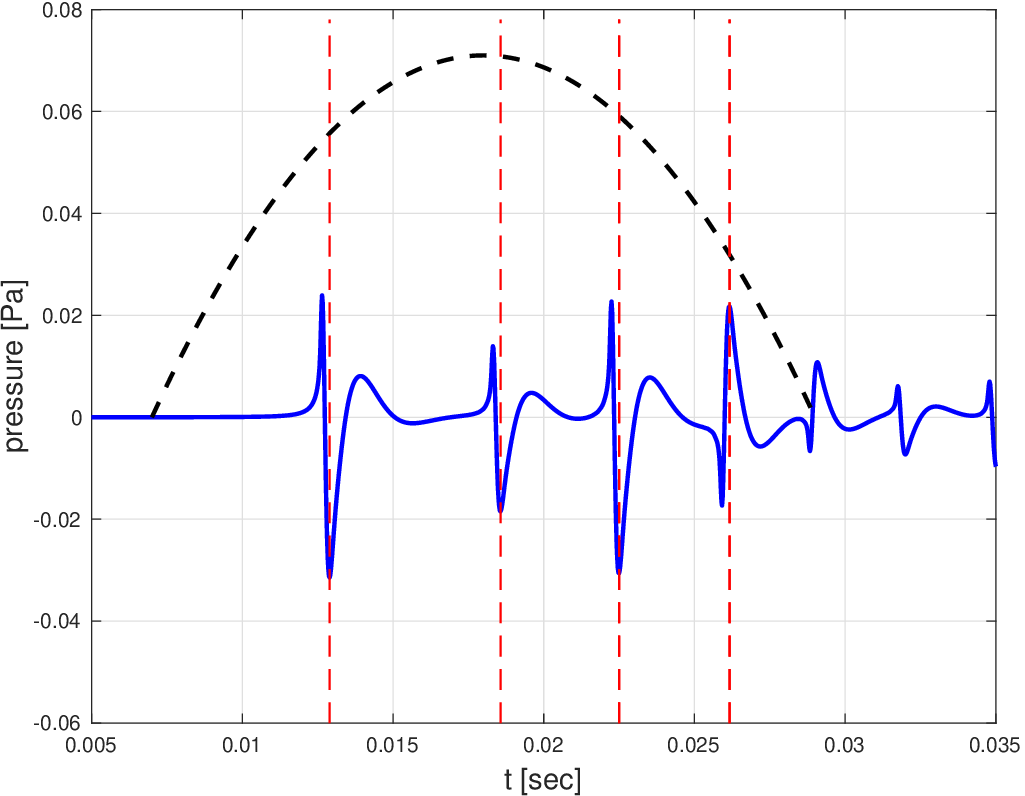}
\caption[(Color online) Simulated RIR of the SH coefficient of order $n=0$ for the spherical microphone array and the spherical loudspeaker array. A Welch window with a 22$\,$ms duration is plotted, starting at $t = 7\,$ms.]{(Color online) Simulated RIR of the SH coefficient of order $n=0$ for the spherical microphone array and the spherical loudspeaker array. A Welch window with a 22$\,$ms duration is plotted, starting at $t = 7\,$ms.}
\label{fig2}
\end{figure}

An example analysis is provided, in which a simulation setup, including room dimensions and spherical loudspeaker array and spherical microphone array positions, is selected so as to generate a scenario with multiple reflections with equal time delays.
A MIMO system is comprised of a spherical loudspeaker array with radius $r_L = 0.1\,$m and SH order $N_L = 3$ and a spherical microphone array with radius $r_M = 0.07\,$m and SH order $N_M = 2$. % frequencies for SLA: 1100    1653    2204 for Nl = 2, 3, 4, and for SMA: 1570 for Nl = 2, giving rm*Nl = 0.2 and rl*Nm = 0.21 (which is pretty close). 
In particular, the loudspeakers' membrane diameter, which is required for calculating $q_n(kr_L)$ from Eq.~\eqref{eq:bck7}, is set to 2-in.
The radii and SH orders of the arrays have been chosen to roughly reflect the orders of currently available spherical arrays$^{28, 29,30}$. 
The system is positioned in a room with dimensions of $(10,\, 10,\, 8)\,$m, for which the reflection coefficients of all walls were set to 0.8 for simplicity. 
The microphone array and loudspeaker array were positioned at $(5,\, 5,\, 3)\,$m and $(2,\, 2,\, 1.75)\,$m, respectively, and multiple transfer functions were simulated using the McRoomSim software \cite{wabnitz2010room} for a sampling frequency of $48\,$kHz.  
The transfer functions were arranged in matrix form, as in Eq.~\eqref{eq:bck4}. 
A general additive error model was employed to model system identification errors, so as to facilitate an analysis that is not constrained to specific system identification methods. 
An error matrix with the same dimensions as the system transfer matrix was simulated and then added to the system room transfer matrix. 
Each of the elements of the error matrix was generated using a zero-mean Gaussian distribution. 
For every frequency, the variance of the errors was set equal for all elements, to produce a $-30\,$dB normalized misalignment projection \cite{zhang2008algorithm} at that frequency. 
Note that while the errors are added to represent practical system identification methods, they can also represent the effect of transducer noise \cite{rafaely2005analysis}. 
An inverse discrete Fourier transform (DFT) was then applied to every element in the system transfer matrix to produce the system RIRs, as an initial step before applying time windowing.  
For reference, the first 35$\,$ms of the RIR corresponding to the SH coefficient of order $n=0$ for both the microphone array and the loudspeaker array is presented in Fig.~\ref{fig2}. 
A time segment within the responses has been selected so as to limit the number of reflections. 
A Welch window of 1056 samples (corresponding to a time duration of 22$\,$ms) was generated, and was positioned in time, starting at $t = 7\,$ms. 
The Welch window was then applied on the inverse DFT of the system and transformed back to the frequency domain. 
The time window is also presented in Fig.~\ref{fig2}.
% former place of Fig. (2)
Six reflections, including the direct sound, are assumed to remain after the application of the time window, whose DORs, DOAs, and time delays, have been calculated analytically and are summarized in Table \ref{table1}.  
% former place of table 1. 
The time delays are also plotted on Fig.~\ref{fig2} as vertical dashed lines. 
In particular, a higher amplitude is seen at $t = 0.0225\,$ms and $t = 0.0262\,$ms, since each of these time samples correspond to two reflections. 
After the application of the time window, the DFT was applied to the system, followed by the application of plane-wave decomposition to the microphone array, as in Eq.~\eqref{eq:bck16}. 
The resulting system is denoted $\bm A(\omega)$. 
Parts, or more specifically, rows, of $\bm A(\omega)$ will be used to define other systems with lower loudspeaker array SH orders, as described in the next section. 

\subsection{Methods}
\justify % after each section and subsection

Eigenvalues are calculated for the smoothed modal cross-spectrum matrices, for modal smoothing, frequency smoothing, and modal smoothing combined with frequency smoothing. 
For modal smoothing, $\bm A(\omega)$ is used for calculating $\tilde{\bm S}_{\bm A}(\omega)$, as in Eq.~\eqref{eq:MS5}, at $f=1.6\,$kHz, which was chosen for analysis as proposed in Ref.~15 %\cite{morgenstern2017design} 
for the given system parameters. 
For frequency smoothing, only a single modal vector, i.e., one column of $\bm A(\omega)$, is required. 
For simplicity, $\bm a_{00}(\omega)$, which corresponds to an omnidirectional directivity at the loudspeaker array, is chosen. 
The smoothed modal cross-spectrum matrix, $\tilde{\bm S}_{\bm a}$, is calculated as in Eq.~\eqref{eq:FS4}, using a frequency band of $[1, 1.6]\,$kHz. 
Note that a small frequency band leads to reflections with long time responses. 
Nonetheless, smoothing is expected to work as long as the number of reflections that contribute to the measured data is limited. 
This may be the case when the early reflections of a RIR, which are relatively sparse, are isolated using time windowing applied to the broadband (and not narrowband) RIRs. 
In addition, a system with a lower loudspeaker array SH order is also investigated to demonstrate the limitations of modal smoothing. 
A system with $N_L = 1$, denoted $\bm A_1(\omega)$, is initially defined by taking the first four columns of $\bm A(\omega)$. $\bm A_1(\omega)$ is used to calculate the smoothed modal cross-spectrum using modal smoothing, $\tilde{\bm S}_{\bm A_1}(\omega)$, as in Eq.~\eqref{eq:MS5}. 
Finally, modal smoothing and frequency smoothing are combined to define a new smoothed modal cross-spectrum matrix for system $\bm A_1(\omega)$. 
The new matrix, denoted $\hat{\bm S}_{\bm A_1}$, is defined similarly to Eq.~\eqref{eq:FS4}, but with averaging over frequency applied to $\tilde{\bm S}_{\bm A_1}(\omega)$, i.e., $\hat{\bm S}_{\bm A_1}$ is defined as: 
\begin{eqnarray}\label{eq:sim1}
\hat{\bm S}_{\bm A_1} & = & \frac{1}{Q} \sum_{q = 1}^{Q} \tilde{\bm S}_{\bm A_1}(\omega_q),  
\end{eqnarray}
with the same frequency bins, or $\omega_q$, as used for $\tilde{\bm S}_{\bm a}$

MUSIC spectra are calculated for some of the smoothed modal cross-spectrum matrices, followed by DOA estimation.  
For modal smoothing, for example, $\tilde{\bm S}_{\bm A}(\omega)$ is initially assumed to have an eigenvalue decomposition of the following form: 
\begin{eqnarray}\label{eq:sim2}
\tilde{\bm S}_{\bm A}(\omega) & = & \bm U \bm \Psi \bm U^{\mathrm{H}},
\end{eqnarray}
where $\bm U$ is a unitary matrix and $ \bm \Psi$ is a diagonal matrix with non-negative eigenvalues on its diagonal. 
The signal subspace is assumed to be spanned by the $L = 6$ eigenvectors that correspond to the largest eigenvalues, and the remaining eigenvectors are assumed to span the noise subspace, denoted by $\bm U_n$. 
Using $\bm U_n$, a MUSIC spectrum, $P_{MUSIC}(\bm \theta)$, is then calculated as: 
\begin{eqnarray}\label{eq:sim3}
P_{MUSIC}(\bm \theta) & = & \frac{1}{\|     \bm U_n  \bm y(\bm \theta) \|^2},
\end{eqnarray}
where $\|  \cdot \|$ is the 2-norm and $\bm y(\bm \theta)$ is the microphone array steering vector, defined as in Eq.~\eqref{eq:bck8s}, but for $N_M$, and spanned over all spatial angles. 
DOAs are then estimated as the angles that correspond to the $L$ maxima of the spectrum.
A MUSIC spectrum is calculated and the corresponding DOAs are estimated for frequency smoothing in a similar manner, but using $\tilde{\bm S}_{\bm a}$ in Eq.~\eqref{eq:sim2}, instead of $\bm S_{A}(\omega)$. 
Note that MUSIC assumes spatially white noise, while the noise is not, in general, spatially white, considering the different processing steps. 
While whitening is typically required for MUSIC, it was found that similar performance is achieved without whitening for RIR data \cite{huleihel2013spherical}, and, therefore, whitening is not applied in this study.  

\subsection{Results}
\justify % after each section and subsection

\begin{figure}[H]
\centering
\includegraphics[clip, trim = {0 0 0 {0 mm}}, width = 1\linewidth]{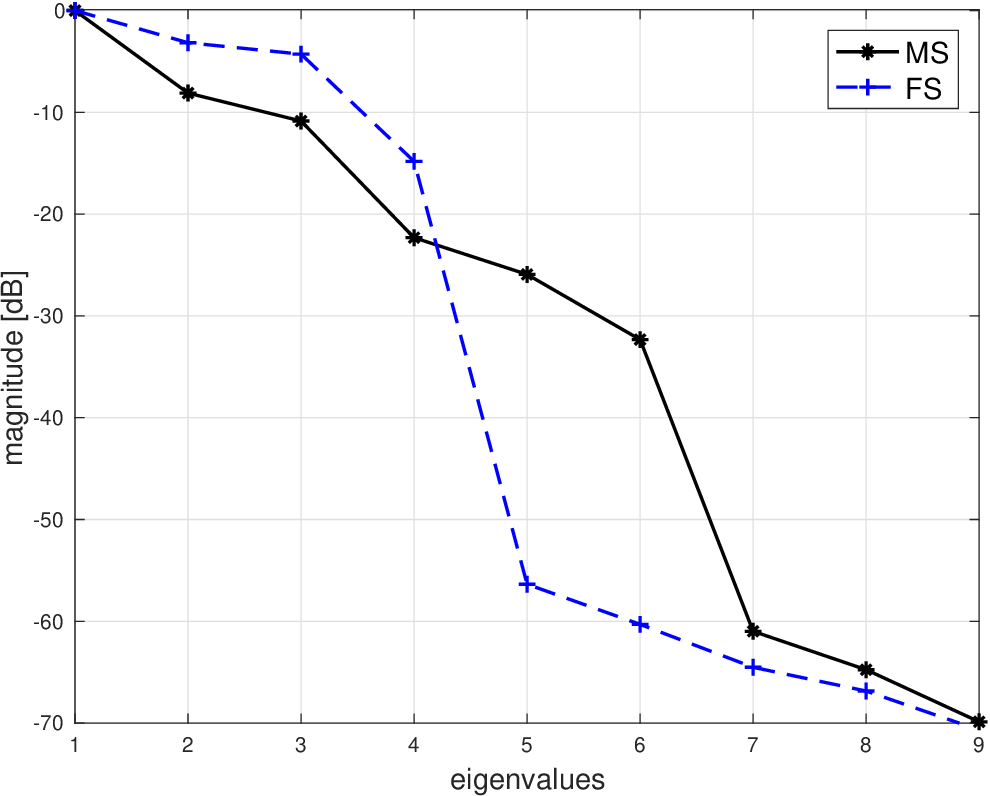}
\caption[(Color online) Eigenvalue distribution of $\tilde{\bm S}_{\bm A}(\omega)$ and $\tilde{\bm S}_{\bm a}$ for modal smoothing (MS) and frequency smoothing (FS), respectively.]{(Color online) Eigenvalue distribution of $\tilde{\bm S}_{\bm A}(\omega)$ and $\tilde{\bm S}_{\bm a}$ for modal smoothing (MS) and frequency smoothing (FS), respectively.}
\label{fig3}
\end{figure}

\begin{figure}[H]
\centering
\includegraphics[clip, trim = {0 0 0 {0 mm}}, width = 1\linewidth]{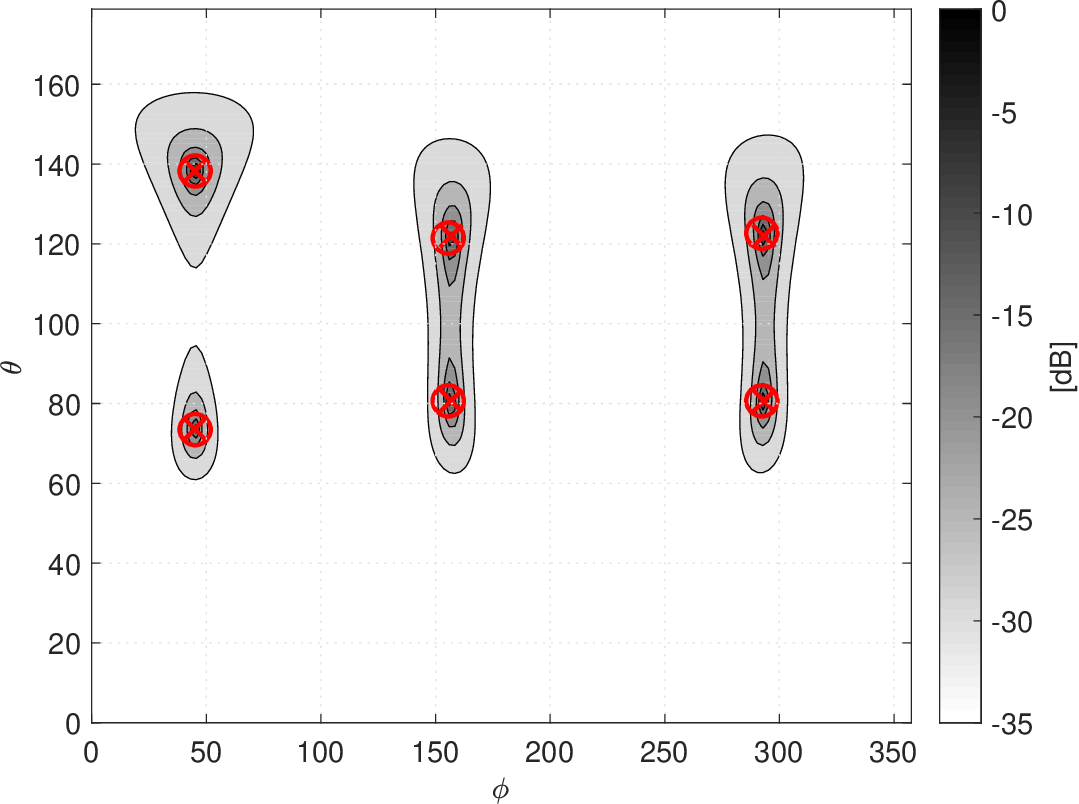}
\caption[(Color online) MUSIC spectrum for modal smoothing, calculated using $\tilde{\bm S}_{\bm A}(\omega)$. `X'-and `O'-marks indicate the true DOAs of the reflections and the estimated DOAs, respectively.]{(Color online) MUSIC spectrum for modal smoothing, calculated using $\tilde{\bm S}_{\bm A}(\omega)$. `X'-and `O'-marks indicate the true DOAs of the reflections and the estimated DOAs, respectively.}
\label{fig4}
\end{figure}

\begin{figure}[H]
\centering
\includegraphics[clip, trim = {0 0 0 {0 mm}}, width = 1\linewidth]{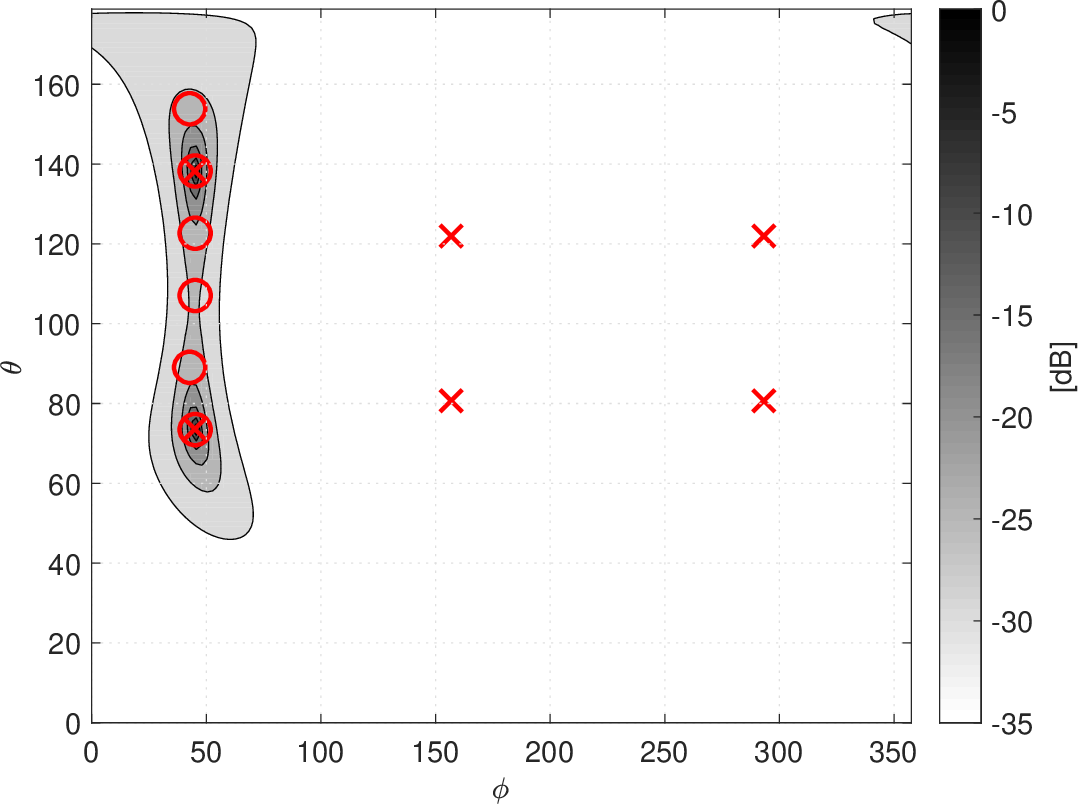}
\caption[(Color online) Same as Fig.~\ref{fig4}, but for frequency smoothing, calculated using $\tilde{\bm S}_{\bm a}$.]{(Color online) Same as Fig.~\ref{fig4}, but for frequency smoothing, calculated using $\tilde{\bm S}_{\bm a}$.}
\label{fig5}
\end{figure}

DOA estimation using MUSIC with modal smoothing or with frequency smoothing is presented for the selected setup. 
In particular, the first step of MUSIC involves a separation into noise and signal subspaces based on the eigenvalue distribution of the modal cross-spectrum matrices. 
The eigenvalue distributions of $\tilde{\bm S}_{\bm a}$ and $\tilde{\bm S}_{\bm A}(\omega)$ for frequency smoothing and modal smoothing, respectively, are presented in Fig.~\ref{fig3}. 
% former position of figure 3
In both cases, two distinct groups of eigenvalues are evident: the signal eigenvalues, with relatively high values (above about -35$\,$dB), and the noise eigenvalues, with relatively low values.  
However, there is a different number of signal eigenvalues for modal smoothing and frequency smoothing: 
six signal eigenvalues are evident for modal smoothing, while only four are evident for frequency smoothing. 
Given $L = 6$, the latter implies that correct separation into signal and noise subspaces is feasible in this case only for modal smoothing. 
This is demonstrated in Figs.~\ref{fig4} and \ref{fig5}, which show the MUSIC spectra for modal smoothing and frequency smoothing, respectively. %, calculated using $\tilde{\bm S}_{A}(k)$ and $\tilde{\bm S}_{\bm a}$ 
% former positions of Figs. 4 and 5
On these figures, `X'-marks indicate the true DOAs of the reflections, calculated analytically given the room geometry and the arrays' positions, and `O'-marks indicate the estimated DOAs, calculated using the MUSIC spectrum. 
Accurate estimations of DOAs are evident in the case of modal smoothing. 
For frequency smoothing, the MUSIC spectrum is distorted, since frequency smoothing fails at decorrelating the reflections with equal time delays, as explained in Sec.~III, %\ref{sec:FS}, 
and peaks in the spectrum are only seen at the angles that correspond to the DOAs of the direct sound and reflection 1 from Table \ref{table1}, which have distinct time delays. 
Therefore, the DOAs of all reflections cannot be estimated using frequency smoothing in this example. 

\begin{figure}[H]
\centering
\includegraphics[clip, trim = {0 0 0 {0 mm}}, width = 1\linewidth]{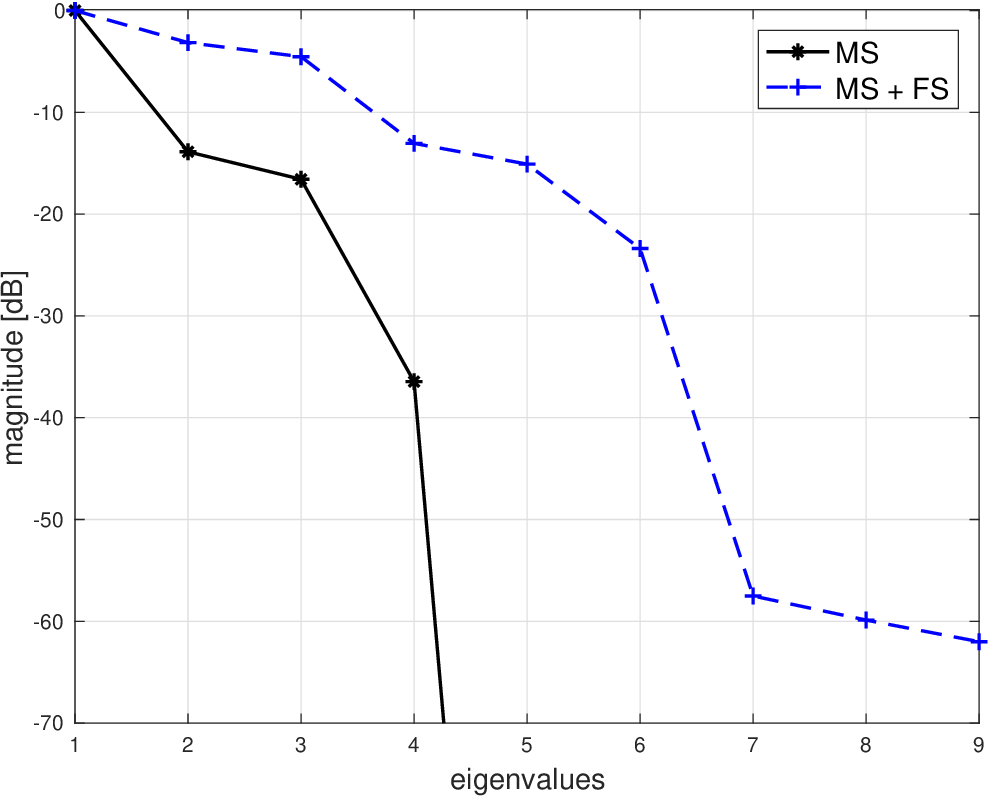}
\caption[(Color online) Eigenvalue distribution of $\tilde{\bm S}_{\bm A_1}$ and $\hat{\bm S}_{\bm A_1}$ for modal smoothing (MS) and modal smoothing combined with frequency smoothing (MS + FS), respectively, for system with $N_L = 1$.]{(Color online) Eigenvalue distribution of $\tilde{\bm S}_{\bm A_1}$ and $\hat{\bm S}_{\bm A_1}$ for modal smoothing (MS) and modal smoothing combined with frequency smoothing (MS + FS), respectively, for system with $N_L = 1$.}
\label{fig6}
\end{figure}

\begin{figure}[H]
\centering
\includegraphics[clip, trim = {0 0 0 {0 mm}}, width = 1\linewidth]{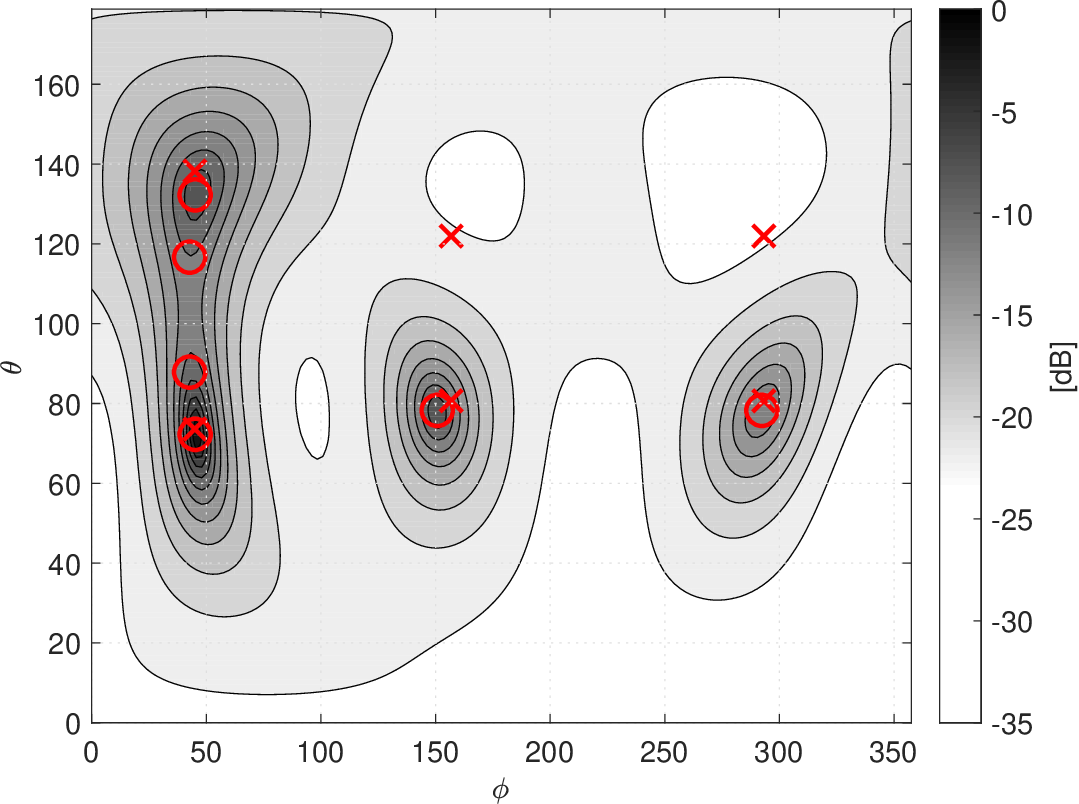}
\caption[(Color online) Same as Fig.~\ref{fig4}, but calculated using $\tilde{\bm S}_{\bm A_1}$ for $N_L=1$. ]{(Color online) Same as Fig.~\ref{fig4}, but calculated using $\tilde{\bm S}_{\bm A_1}$ for $N_L=1$. }
\label{fig7}
\end{figure}

\begin{figure}[H]
\centering
\includegraphics[clip, trim = {0 0 0 {0 mm}}, width = 1\linewidth]{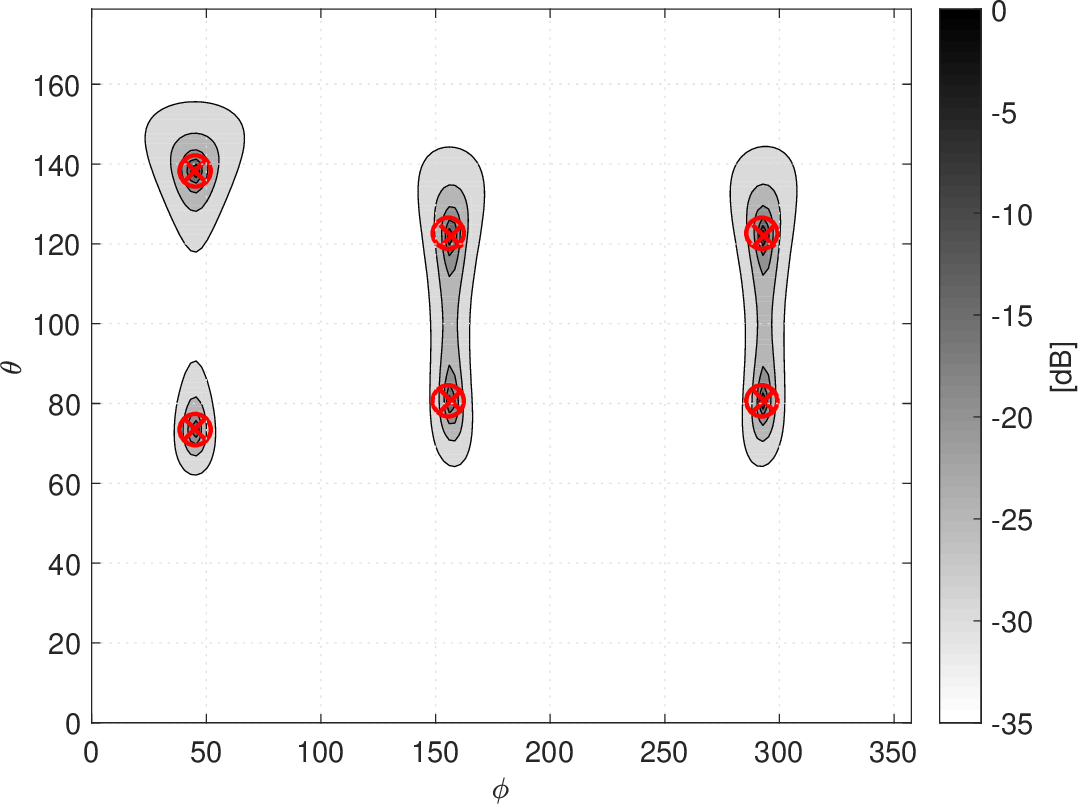}
\caption[(Color online) Same as Fig.~\ref{fig4}, but for modal smoothing combined with frequency smoothing, calculated using $\hat{\bm S}_{\bm A_1}$ for $N_L=1$. ]{(Color online) Same as Fig.~\ref{fig4}, but for modal smoothing combined with frequency smoothing, calculated using $\hat{\bm S}_{\bm A_1}$ for $N_L=1$. }
\label{fig8}
\end{figure}

Modal smoothing and frequency smoothing can be combined for DOA estimation for MIMO systems with low loudspeaker array SH orders. 
Low SH orders impose a limitation on the number of DOAs that can be estimated using modal smoothing, as discussed in Sec.~IV. %\ref{sec:MS}. 
While frequency smoothing by itself was also shown to fail in the setup under study, modal smoothing combined with frequency smoothing can be employed to increase the number of reflections that can be estimated.  
This is demonstrated for a system with $N_L  = 1$ in Fig.~\ref{fig6}, which presents the eigenvalue distributions of $\tilde{\bm S}_{\bm A_1}$ and $\hat{\bm S}_{\bm A_1}$ for modal smoothing and for modal smoothing combined with frequency smoothing, respectively. 
% former position of Fig. 6
A distinct separation between the signal eigenvalues and the noise eigenvalues is seen for both cases in the figure, as in Fig.~\ref{fig3}.
Only four signal eigenvalues are seen for modal smoothing, due to the use of the lower SH order of the loudspeaker array, which employs averaging over only four channels. 
For modal smoothing combined with frequency smoothing, six signal eigenvalues are seen. 
This implies that correct separation into signal and noise subspaces is only feasible for modal smoothing combined with frequency smoothing in this case.   
This is further demonstrated in Figs.~\ref{fig7} and \ref{fig8}, which show the MUSIC spectra for modal smoothing and modal smoothing combined with frequency smoothing.
% former position of Figs 7 and 8
In conclusion, this simulation study demonstrated the successful applicability of modal smoothing for DOA estimation of room reflections with equal time delays. 
This has been demonstrated for a specific configuration;
however, the results can be generalized for other SH orders of the arrays, frequencies, and powers of errors following the analysis in Ref.~14. %\cite{morgenstern2015theory}. 
Finally, combining modal smoothing with frequency smoothing has been shown to be beneficial in the case of insufficient loudspeaker array SH orders relative to the number of reflections. 
In this case, the loudspeaker array SH order imposes a limitation on the number of reflections with equal time delays, and not on the number of reflections in total when only modal smoothing is applied. 
Continuing with the simulation example of the system with $N_L = 1$, modal smoothing fails in this case, as previously discussed, since $(N_L+1)^2$ is smaller than six, which is the number of reflections including the direct sound. 
However, modal smoothing combined with frequency smoothing performs well in this case; 
modal smoothing is able to decorrelate each set of reflections with equal time delays, since the maximum number of reflections that have equal time delay is two, which is less than $(N_L+1)^2$.  
Frequency smoothing is able to decorrelate reflections with different time delays, with the total number of reflections that can be estimated limited to  $(N_M+1)^2-1$ \cite{khaykin2012acoustic}.

\section{Conclusions}\label{sec:conclusions}
\justify % after each section and subsection
 
Modal smoothing for MIMO systems was proposed for restoring the rank of cross-spectrum matrices in the case of room reflections with equal time delays. 
Results show that modal smoothing achieves full rank restoration in this case, which leads to accurate DOA estimations using MUSIC, whereas frequency smoothing, an example of previous methods, fails. 
Combining modal smoothing with frequency smoothing was also shown to achieve rank restoration, which implies that simple systems with spherical loudspeaker arrays of low SH orders can be employed in practice where different reflections reach the microphone array at the same time. 

\section{ACKNOWLEDGMENTS}\label{sec:ACK}
\justify % after each section and subsection

This research was supported by The Israel Science Foundation (Grant No.~146/13). 

%\end{linenumbers}
%\medskip

\newpage
\justify
\begin{table}
\centering
\begin{tabular}{| c | c | c | c|}
 \hline
reflection & time delay [sec]  & DOR $(\beta, \psi)$ [deg] & DOA $( \theta, \phi)$ [deg]  \\ \hline \hline
Direct sound &0.0129 & (106.4, 225) & (73.6, 45) \\ \hline
1 & 0.0186 & (138.2, 225) & (138.2, 45) \\ \hline
2 & 0.0225 & (99.3, 246.8) & (80.7, 293.2) \\ \hline
3 & 0.0225 & (99.3, 203.2) & (80.7, 156.8) \\ \hline
4 & 0.0262 & (122, 246.8) & (122, 293.2) \\ \hline
5 & 0.0262 & (122, 203.2) & (122, 156.8) \\ \hline
\end{tabular}
\caption{Time delays, DORs, and DOAs of the direct sound and the first five reflections in the RIR presented in Fig.~\ref{fig2}.}
\label{table1}
\end{table}

% figures: 
%\newpage
%\renewcommand{\listfigurename}{List of figures} 
%\listoffigures



\begin{thebibliography}{10}
% 1, 

\bibitem{ribeiro2010using}
F.~Riveiro, C.~Zhang, D.A.~Flor{\^e}ncio, D.E.~Ba, ``Using reverberation to improve range and elevation discrimination for small array sound source localization,'' IEEE
  Transactions on Acoustics, Speech, and Signal Processing 18~(7) (2010)
  1781--1792.
  
  \bibitem{antonacci2010geometric}
F.~Antonacci, A.~Sarti, S.Tubar, 
J.~Meyer, G.~Elko, ``Geometric reconstruction of the environment
 from its response to multiple acoustic emissions,'' in: Acoustics, Speech, and
  Signal Processing (ICASSP), 2010 IEEE International Conference on, Vol.~2,
  IEEE, pp. 2822--2825 (Dallas, Texas, USA, 2010).

\bibitem{gover2004measurements}
B.N.~Gover, J.G.~Ryan, M.R.~Stinson, ``Measurements of directional properties
 of reverberant sound fields in rooms using a spherical microphone array,'' The Journal of the Acoustical Society
  of America 116~(4) (2004) 2138--2148.
  
  \bibitem{khaykin2012acoustic}
D.~Khaykin, B.~Rafaely, ``Acoustic analysis by spherical microphone array
  processing of room impulse responses'', The Journal of the Acoustical Society
  of America 132~(1) (2012) 261--270.
  
  \bibitem{kuster2004acoustic}
  M.~Kuster, D.~de Vries, E.~Hulsebos, A.~Gisolf, ``Acoustic imaging in
   enclosed spaces: Analysis of room geometry modifications on the impulse response,'' The Journal of the Acoustical Society
  of America 116~(4) (2004) 2126--2137.

\bibitem{gover2002microphone}
B.N.~Gover, J.G.~Ryan, M.R.~Stinson, ``Microphone array measurement
 system for analysis of directional and spatial variations of sound fields,'' The Journal of the Acoustical Society
  of America 112~(5) (2002) 1980--1991.


\bibitem{van2004detection}
H.~L. Van~Trees, ``Detection, estimation, and modulation theory, optimum array
  processing,'' (John Wiley \& Sons, Inc., New York, 2002), pp.1158--1163, 1233--1242.  
  
  \bibitem{yan2011optimal}
  S.~Yan, H.~Sun, U.P~Svensson, X.~Ma, J.M.~Hovem, ``Optimal modal beamforming for spherical microphone arrays,'' 
   IEEE Transactions on Signal Processing 19~(2) (2011) 361--371.
  
\bibitem{sun2011joint}
H.~Sun, H.~Teutsch, E.~Mabande, W.~Kellermann, ``Robust localization of multiple sources in reverberant environments using EB-ESPRIT with spherical microphone arrays,'' in: Acoustics, Speech, and
  Signal Processing (ICASSP), 2011 IEEE International Conference on, 
  IEEE, pp. 117--120 (Prague, Czech Republic, 2011).

\bibitem{sun2012localization}
H.~Sun, E.~Mabande, K.~Kowalczyk, W.~Kellermann, ``Localization of distinct reflections in rooms using spherical microphone array eigenbeam processing,'' The Journal of the Acoustical Society
  of America 131~(4) (2012) 2828--2840.

\bibitem{wang1985coherent}
H.~Wang, M.~Kaveh, ``Coherent signal-subspace processing for the detection and
  estimation of angles of arrival of multiple wide-band sources,'' IEEE
  Transactions on Acoustics, Speech, and Signal Processing 33~(4) (1985)
  823--831.

\bibitem{doron1994wavefield}
M.~A. Doron, E.~Doron, ``Wavefield modeling and array processing. ii. algorithms,''
  IEEE Transactions on Signal Processing 42~(10) (1994) 2560--2570.

\bibitem{huleihel2013spherical}
N.~Huleihel, B.~Rafaely, ``Spherical array processing for acoustic analysis using
  room impulse responses and time-domain smoothing,'' The Journal of the
  Acoustical Society of America 133~(6) (2013) 3995--4007.

\bibitem{morgenstern2015theory}
H.~Morgenstern, B.~Rafaely, F.~Zotter, ``Theory and investigation of acoustic
  multiple-input multiple-output systems based on spherical arrays in a room,''
  The Journal of the Acoustical Society of America 138~(5) (2015) 2998--3009.

\bibitem{morgenstern2017design}
H.~Morgenstern, B.~Rafaely, M.~Noisternig, ``Design framework for spherical
  microphone and loudspeaker arrays in a multiple-input multiple-output system,''
  The Journal of the Acoustical Society of America 141~(3) (2017) 2024--2038.

\bibitem{morgenstern2017spatial}
H.~Morgenstern, B.~Rafaely, ``Spatial reverberation and dereverberation using an
  acoustic multiple-input multiple-output system,'' Journal of the Audio
  Engineering Society 65~(1/2) (2017) 42--55.

\bibitem{morgenstern2013enhanced}
H.~Morgenstern, B.~Rafaely, ``Enhanced spatial analysis of room acoustics using
  acoustic multiple-input multiple-output (mimo) systems,'' in: Proceedings of
  Meetings on Acoustics, Vol.~19, no.~1, (Acoustical Society of America, Montreal, Canada, 2013), pp.~15--18. 
  
 \bibitem{rafaely2004plane}
B.~Rafaely, ``Plane-wave decomposition of the sound field on a sphere by
  spherical convolution,'' The Journal of the Acoustical Society of America 116
  (2004) 2149.


\bibitem{morgenstern2014farfield}
H.~Morgenstern, B.~Rafaely, ``Far-field criterion for spherical microphone arrays
  and directional sources,'' in: Hands-free Speech Communication and Microphone
  Arrays (HSCMA), 2014 Joint Workshop on, pp.~42--46 (Nancy, France, 2014).
  
\bibitem{driscoll1994computing}
J.~R. Driscoll, D.~M. Healy, ``Computing Fourier transforms and convolutions on
  the 2-sphere,'' Advances in applied mathematics 15~(2) (1994) 202--250.

\bibitem{williams1999fourier}
E.~Williams, ``Fourier acoustics: sound radiation and nearfield acoustical
  holography,'' chapters 2.4, 6.7.10 (Academic Press, London, 1999).

\bibitem{rafaely2011optimal}
B.~Rafaely, D.~Khaykin, ``Optimal model-based beamforming and 
independent steering for spherical loudspeaker arrays,'' IEEE Trans.~Audio, Speech, Lang.~Process. 19(7), 2234–2238 (2011).

\bibitem{tourbabin2015consistent}
V.~Tourbabin, B.~Rafaely, ``On the consistent use of space and time conventions
  in array processing,'' Acta Acustica united with Acustica 101~(3) (2015)
  470--473.
  
\bibitem{rafaely2009spherical}
B.~Rafaely, ``Spherical loudspeaker array for local active control of sound,'' The
  Journal of the Acoustical Society of America 125~(5) (2009) 3006--3017.

\bibitem{allen1979image}
J.~B. Allen, D.~A. Berkley, ``Image method for efficiently simulating small-room
  acoustics,'' The Journal of the Acoustical Society of America 65 (1979) 943.

\bibitem{rafaely2005analysis}
B.~Rafaely, ``Analysis and design of spherical microphone arrays, Speech and
  Audio Processing,'' IEEE Transactions on 13~(1) (2005) 135--143.

\bibitem{roy2007effective}
O.~Roy, M.~Vetterli, ``The effective rank: A measure of effective dimensionality,''
  in: 15th European Signal Processing Conference (EUSIPCO), IEEE, 2007 pp.~606--610 (Poznan, Poland, 2007).

\bibitem{meyer2002highly}
J.~Meyer, G.~Elko, ``A highly scalable spherical microphone array based on an
  orthonormal decomposition of the soundfield,'' in: Acoustics, Speech, and
  Signal Processing (ICASSP), 2002 IEEE International Conference on, Vol.~2,
  IEEE, pp. II--1781 (Orlando, Florida, USA, 2002).
  
\bibitem{pasqual2010theoretical}
A.~M.~Pasqual, P.~Herzong, J.~R.~Arruda, ``Theoretical and experimental analysis of the electromechanical behavior of a compact spherical loudspeaker array for directivity control,'' The Journal of the Acoustical Society of America 128~(6) (2010) 3478--3488.

\bibitem{Klein2014optimized}
J.~Klein, M.~Pollow, M.~Vorlaender, ``Optimized Spherical Sound Source for Auralization with Arbitrary Source Directivity,'' in: Proc. EAA Joint Symposium on Auralization and Ambisonics, pp.~56--61 (Berlin, Germany, 2014).

\bibitem{wabnitz2010room}
A.~Wabnitz, N.~Epain, C.~Jin, A.~van Schaik, ``Room acoustics simulation for
  multichannel microphone arrays,'' in: Proceedings of the International
  Symposium on Room Acoustics, pp.~1--6 (Melbourne, Australia, 2010).

\bibitem{zhang2008algorithm}
W.~Zhang, P.~A. Naylor, ``An algorithm to generate representations of system
  identification errors,'' Journal of Electrical and Computer Engineering, Research letters in signal processing (Hindawi Publishing Corp., Cairo, Egypt, 2008), 13.
 
\end{thebibliography}
\end{document}